\documentclass[aps,prd,preprint,superscriptaddress,tightenlines,nofootinbib]{revtex4}

\usepackage{graphicx}
\usepackage{dcolumn}
\usepackage{bm}

\begin{document}

\preprint{CLNS 06/1952}       
\preprint{CLEO 06-02}        

\newcommand{\ccbar}{c \bar c}
\newcommand{\piz}{\pi^0}
\newcommand{\etaprime}{\eta\,'}
\newcommand{\diel}{e^+e^-}
\newcommand{\dimu}{\mu^+\mu^-}
\newcommand{\dika}{K^+K^-}
\newcommand{\dipi}{\pi^+\pi^-}
\newcommand{\dipiz}{\pi^0\pi^0}
\newcommand{\ThreePi}{\pi^+\pi^-\piz}
\newcommand{\dilep}{{\ell^+\ell^-}}
\newcommand{\KK}{K^+K^-}
\newcommand{\jpsi}{{J/\psi}}
\newcommand{\psip}{{\psi(2S)}}
\newcommand{\pipijpsi}{\pi^+\pi^-\jpsi}
\newcommand{\pizpizjpsi}{\pi^0\pi^0\jpsi}
\newcommand{\kkjpsi}{K^+K^-\jpsi}
\newcommand{\etajpsi}{\eta\jpsi}
\newcommand{\pizjpsi}{\piz\jpsi}
\newcommand{\etaetajpsi}{\eta\eta\jpsi}
\newcommand{\etaprimejpsi}{\etaprime\jpsi}
\newcommand{\threepijpsi}{\ThreePi\jpsi}
\newcommand{\pipipsiprime}{\pi^+\pi^-\psi(2S)}
\newcommand{\etapsiprime}{\eta\psi(2S)}
\newcommand{\pipiphi}{\pi^+\pi^-\phi}
\newcommand{\chicJ}{\chi_{cJ}}
\newcommand{\gammachicone}{\gamma\chi_{c1}}
\newcommand{\gammachictwo}{\gamma\chi_{c2}}
\newcommand{\threepichicone}{\ThreePi\chi_{c1}}
\newcommand{\threepichictwo}{\ThreePi\chi_{c2}}
\newcommand{\omegachiczero}{\omega\chi_{c0}}
\newcommand{\gev}{\,\mbox{GeV}}
\newcommand{\mev}{\,\mbox{MeV}}
\newcommand{\ev}{\,\mbox{eV}}
\newcommand{\pb}{\,\mbox{pb}}
\newcommand{\invpb}{\,\mbox{pb}^{-1}}
\newcommand{\yfts}{Y(4260)}
\newcommand{\psiff}{\psi(4040)}
\newcommand{\psifos}{\psi(4160)}

\title{Charmonium Decays of $Y(4260)$, $\psi(4160)$, and $\psi(4040)$}

\author{T.~E.~Coan}
\author{Y.~S.~Gao}
\author{F.~Liu}
\affiliation{Southern Methodist University, Dallas, Texas 75275}
\author{M.~Artuso}
\author{S.~Blusk}
\author{J.~Butt}
\author{J.~Li}
\author{N.~Menaa}
\author{R.~Mountain}
\author{S.~Nisar}
\author{K.~Randrianarivony}
\author{R.~Redjimi}
\author{R.~Sia}
\author{T.~Skwarnicki}
\author{S.~Stone}
\author{J.~C.~Wang}
\author{K.~Zhang}
\affiliation{Syracuse University, Syracuse, New York 13244}
% ...
\collaboration{CLEO Collaboration} 
\noaffiliation

\date{February 20, 2006}

\begin{abstract} 
Using data collected with the CLEO detector operating at the 
CESR $\diel$ collider at $\sqrt s = 3.97$-$4.26\gev$,
we investigate 15~charmonium decay modes of the
$\psiff$, $\psifos$, and $\yfts$ resonances. We confirm,
at 11$\sigma$ significance,
the {\sc BaBar} $\yfts \to \pi^+\pi^-\jpsi$ discovery, 
make the first observation of 
$\yfts \to \pizpizjpsi$ ($5.1 \sigma$), and find the first 
evidence for $\yfts \to \kkjpsi$ ($3.7 \sigma$). 
We measure $\diel$ cross-sections at $\sqrt{s}=4.26\gev$ 
as $\sigma(\pipijpsi)$=$58^{+12}_{-10}$$\pm 4 \pb$,
$\sigma(\dipiz\jpsi)$=$23^{+12}_{-8}$$\pm1 \pb$, and
$\sigma(\dika\jpsi)$=$9^{+9}_{-5}$$\pm1 \pb$, in which the
uncertainties are statistical and systematic, respectively.
Upper limits are placed on other decay rates
from all three resonances.
\end{abstract}

\maketitle

The region at center-of-mass energies
above charmonium open-flavor production threshold
is of great interest to theory due to its richness 
of $c\bar{c}$ states, the properties of which are
not well-understood. Prominent structures
in the hadronic cross-section are the 
$\psi(3770)$, the $\psi(4040)$, and the $\psi(4160)$~\cite{PDG}.
Their main characteristics are large total widths,
two orders of magnitude larger than for the
lower-lying $\ccbar$~states of $J^{PC}=1^{--}$, and 
weaker couplings to leptons than the $\jpsi$ and $\psi(2S)$.
Decays to closed-charm final states are not favored 
due to the availability of open-charm channels.

  Recently, observations of new charmonium-like states in
the same energy region 
have been reported~\cite{homelessmesons}. 
Most of these have been observed to decay through open charm.
However, an enhancement in the invariant mass spectrum of
the closed-charm $\pipijpsi$~final state 
has also been observed by {\sc BaBar}
in initial state radiation (ISR),
$e^+e^- \to \gamma (\pipijpsi)$~\cite{BaBar4260ISR}.
A weaker signal was observed in
the decay $B \to K (\pipijpsi)$~\cite{BaBar4260BDecay}. 
The observed lineshape can be described by a single 
resonance, termed $Y(4260)$, of mass $M = 4259\pm8^{+2}_{-6} \mev$, 
width $\Gamma_{\mathrm{tot}} = 88\pm23^{+6}_{-4} \mev$, and
coupling 
$\Gamma_{ee}\times{\cal B}(Y(4260)\to\dipi\jpsi)
=5.5\pm1.0^{+0.8}_{-0.7}\ev$. It is located 
quite unexpectedly at a local {\it minimum} of the hadronic cross-section.
Since it is observed in ISR, the new state must have 
$J^{PC} = 1^{--}$, and
therefore it can be studied directly in $e^+e^-$ collisions at threshold.
No other evidence for a resonance at this mass
has been identified~\cite{BaBarppbar}, leaving the existence 
and possible charmonium-like nature of this 
state
uncertain.
Many interpretations have been 
suggested; 
to be compatible with the absence of a corresponding
enhancement in open charm production, 
most favor an unconventional explanation of $\yfts$,
such as hybrid charmonium~\cite{ClosePage,hybrids}, 
tetraquarks~\cite{Ebert,Maiani,TWQCD}, 
or hadronic molecules~\cite{LZL,Qiao,WMY}.
One proposal~\cite{Estrada} argues that $Y(4260)$ is conventional:
It identifies $\yfts$ with the $\psi(4S)$ vector $\ccbar$~state,
and relies upon interference effects
to produce the dip in open-charm cross-section
and a hypothesized large coupling to $\dipi\jpsi$ of the $\psi(3S)$,
commonly associated with the $\psi(4040)$.

\begin{figure}
\includegraphics*[width=5.0in]{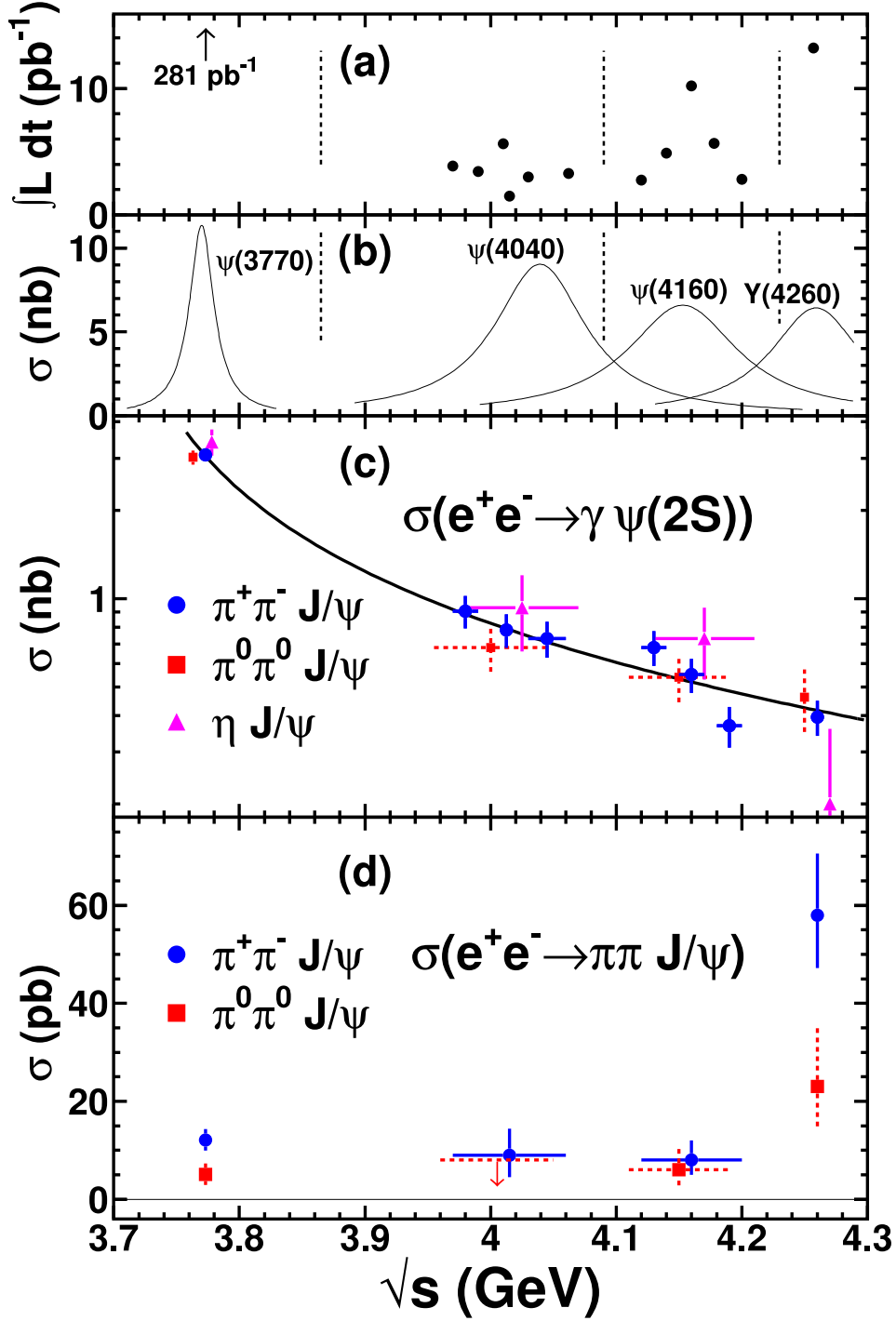}
\caption{(a) $\int{\cal L}dt$ {\sl vs.}~$\sqrt{s}$ (circles). 
(b) Born-level Breit-Wigner cross-sections for resonances in this
energy region; the $\yfts$ curve has an arbitrary vertical scale.
(c) $\diel\to\gamma\psip$ cross-section measurements 
{\sl vs.}~$\sqrt{s}$ from
$\dipi\jpsi$ (circles), $\dipiz\jpsi$ (squares, dashed lines), and $\eta\jpsi$
(triangles) overlaid with the theoretical prediction as described
in the text. 
(d) Cross-sections for $\diel\to\dipi\jpsi$ (circles)
and $\dipiz\jpsi$ (squares, dashed lines) {\sl vs.}~$\sqrt{s}$.
Some points in (c) and (d)
are offset in $\sqrt s$ by 10$\mev$ for display purposes.}
\label{fig:fig1}
\end{figure}

To further clarify the nature of $Y(4260)$, investigation of 
both open and closed charm is necessary.
Here, we report production cross-sections measurements
of 16~final states containing a 
$\jpsi$, $\psi(2S)$, $\chicJ$, or $\phi$,
in the $\psi(4040)$, $\psi(4160)$, and $Y(4260)$ energy region,
motivated by the range of experimental tests suggested so 
far~\cite{ClosePage,hybrids,Ebert,Maiani,TWQCD,LZL,Qiao,WMY,Estrada}.
We use data taken during a scan of center-of-mass energies
$\sqrt s =$3.97-4.26$\gev$, complementing 
a sample 
of an integrated luminosity $\int {\cal L} dt = 281\invpb$ 
at $\sqrt s = 3.773\gev$ 
previously acquired.
Collisions were registered with the CLEO detector~\cite{CLEO} at the 
CESR $\diel$~collider~\cite{CESR}. 
The scan data naturally separate into three regions:
the $\psi(4040)$ ($\sqrt s = 3.97$-$4.06\gev$, 
$ \int {\cal L} dt = 20.7\invpb$),
the $\psi(4160)$ (4.12-4.20$\gev$,
26.3$\invpb$),
and $\sqrt s = 4.26\gev$ (13.2$\invpb$).
Fig.~\ref{fig:fig1}(a) shows a profile of $\int{\cal L}dt$ {\sl vs.}~$\sqrt{s}$
and Fig.~\ref{fig:fig1}(b) the Born-level
Breit-Wigner lineshapes~\cite{PDG}
for the four resonances between $\sqrt{s}$=3.7 and 4.4~GeV,
also indicating the grouping of scan points.
We estimate the number of resonances produced 
by folding together the luminosities with the resonance 
Breit-Wigner cross-sections~\cite{NewRFit},
including radiative corrections~\cite{ISR,RadiativeCorrections}, 
and arrive at
(93$\pm$11)$\times 10^3$ [(115$\pm$15)$\times 10^3$] 
$\psi(4040)$ [$\psi(4160)$] mesons,
where the dominant errors are the uncertainties on
resonance parameters.

The CLEO detector~\cite{CLEO} features a solid angle coverage of $93\%$ for
charged and neutral particles.
The charged particle tracking system operates in a 1.0~T~magnetic field
parallel to the beam axis and achieves a momentum resolution of
$\sim$0.6\% at momenta of $1\gev/c$. The CsI crystal
calorimeter attains
photon energy resolutions of $2.2\%$ for $E_\gamma$=$1\gev$
and $5\%$ at $100\mev$. 
Particle identification is performed with the specific
ionization loss ($dE/dx$) and the Ring Imaging Cherenkov detector (RICH).
Muons of momentum $p>1\gev$ are separated from pions 
by their penetration of the calorimeter, solenoid coil,
and up to three 36-cm-thick slabs of magnet iron
for subsequent detection by wire chambers behind each slab.
The integrated luminosity was measured 
using $\diel$, $\gamma\gamma$, and $\dimu$
events~\cite{LUMINS,Babayaga}. 

The final states analyzed here are listed in Table~\ref{tab:yields}.
We require all particles in each final state to be reconstructed, and
four-momentum conservation is enforced. 
  Mass windows for the following
light hadron decays are set, based on MC studies:
$\piz\to\gamma\gamma$ (110-150$\mev$);
$\eta \to \gamma\gamma$, $\eta \to \ThreePi$ (450-650$\mev$);
$\etaprime\to \dipi\eta$, $\etaprime \to \gamma\dipi$ (930-980$\mev$); 
$\omega \to \ThreePi$ (730-830$\mev$); and
$\phi\to K^+K^-$ (1.00-1.04$\gev$). 

We identify a $\jpsi$ or $\psip$ through
its decay into $\dilep$, $\dilep$=$e^+e^-$ or $\mu^+\mu^-$.
A lepton candidate is 
identified by
the ratio of the energy deposited in the
calorimeter, $E$, to the measured momentum, $p$. 
Muon pair candidates must satisfy $E/p<$0.25 
for at least one of the tracks and $E/p<$0.5 for the other;
both electrons must have $E/p>$0.85 and 
a $dE/dx$ consistent with the value expected for an electron. 
Pions faking muons are additionally suppressed
by requiring at least one muon candidate per
muon pair to leave a signature in the muon system.
A lepton pair is classified as a $\jpsi$ $[\psi(2S)] \to\dilep$ 
candidate if the invariant mass of the decay
products lies within 3.04-3.14 [3.64-3.73]$\gev$.
We also use the decay $\psi(2S) \to \dipi\jpsi$, 
where we require $M(\dipi\jpsi)=$3.64-3.73$\gev$. 

A $\gamma\chi_{c1,2}\to\gamma\gamma\jpsi$  decay is tagged 
by the highest energy photon in the event 
having an energy within ($-$60,$+$40)~MeV 
of $E_\gamma$=$(s-M_{\chi_{cJ}}^2)/2\sqrt{s}$.
Events with $M({\gamma\gamma})$
in the $\piz$ or $\eta$ mass ranges are excluded. 
Any calorimeter shower other than the two 
radiative transition photon candidates
must have $E<$50$\mev$.
We select $\ThreePi\chi_{cJ}$, $\chi_{cJ} \to
\gamma\jpsi$ events ($J=1,2$) by demanding that 
$M(\gamma\jpsi)$ as well as the mass recoiling against the $\ThreePi$
match $M_{\chi_{cJ}}$~\cite{PDG} within $\pm 20\mev$. 

For $\dika\jpsi$, the kaons have momenta that 
are too low for use of the RICH detector, so $\pi^\pm$ rejection
is achieved by requiring that at least one of
the $K^\pm$ candidates have momentum in the range 0.2-0.5$\gev/c$
and have $dE/dx$ within three
standard deviations of the expected value for a $K^\pm$.
For $\pipiphi$ and $\chi_{c0} \to 2(\dipi)$ or $\dipi\dika$, 
$K^\pm$ candidates must be positively
identified as a kaon using a likelihood based upon
both $dE/dx$ and RICH responses, but none of the $\pi^\pm$ candidates
in either decay can be so identified as kaons.
For the $\omega\chi_{c0}$ mode, both the mass of the
$\chi_{c0}$ decay products and the
mass recoiling against the $\omega$
must lie within 50~MeV of $M(\chi_{c0})$~\cite{PDG}.
Similarly, for $\dipi\phi$, the mass recoiling against
$\dipi$ must lie in the range 0.94-1.10$\gev$.

For $\dipi\jpsi$ and $\dipi\psip$,
we require that $M(\dipi)>350\mev$ and
that neither pion candidate be 
identified as an electron via $E/p$ and $dE/dx$ as above. These cuts 
suppress $e^+e^-\to\ell^+\ell^-\gamma$, $\gamma\to e^+e^-$ events
in which the $e^+e^-$ pair from the photon conversion
is mistaken for the $\pi^+\pi^-$.
For $\pi^0 J/\psi$ as well as 
the $\eta J/\psi$
and $\gamma\chi_{c1,2}$ modes
that end in $\gamma\gamma\diel$, 
background from Bhabha events 
is diminished by requiring $\cos\theta_{e^+}<0.3$.

For modes with a $\jpsi$ or $\psip$
we restrict the missing momentum~$k$, computed from the measured event
momenta according to Eq.~(6) of Ref.~\cite{gammaeePsi2S}, 
to further reduce background.
Signal events will have $k\approx0$; selection windows
are set separately for each mode according to the 
measurement resolution predicted
by MC simulation, and range from $\pm 10\mev$ ($\dika\jpsi$)
and $\pm$15$\mev$ ($\dipi\jpsi$) to ($-$35,$+$75)$\mev$ ($\dipiz\jpsi$).

Backgrounds in the $X\jpsi$ 
modes from $\diel\to\gamma^*$$\to\,$hadrons
are estimated from the yields seen in the data with
the restrictions on $M(\dilep)$ and $k$ adjusted to correspond
to masses either below or above $M_\jpsi$.
The sideband windows, $|k|<150\mev$ and either
$M(\dilep)$=2.90$\pm$0.15$\gev$ or 3.30$\pm$0.15$\gev$,
are considerably wider than the signal regions, 
allowing for the accumulation of more
statistics in the background estimates, and therefore are scaled
down by the sideband/signal window width ratios 
prior to subtraction from the
signal yields. 
The sideband scale factors
are verified in MC~simulations of $\diel \to\gamma^* \to\,$hadrons.
We perform a mass sideband subtraction for 
the $\chi_{c0}$, 
as a result of which all observed events can be attributed to
background, and for~$\phi$ and $\etaprime$, 
also reducing the yield substantially; photon energy
sidebands are used to estimate background in $\gamma\chicJ$. 
Background in $\pi\pi\jpsi$, $\eta\jpsi$, and $\gamma\chicJ$
modes produced in $\gamma\psi(2S)$ decay (with 
$M(\psi(2S)) \approx \sqrt s \gg 3686\mev$) is summed along with that
from the sidebands in Table~\ref{tab:yields}, and
comprises 10-40\% of the total background in those modes.
Other modes could
be subjected to similar subtractions, but due to the absence
of significant event populations, we forgo this option,
and use uncorrected yields to compute (conservative) upper limits.

The {\sc evtgen} event generator~\cite{EvtGen}, which includes
final state radiation~\cite{PHOTOS}, and a {\sc geant}-based~\cite{GEANT} 
detector simulation are used to model the physics processes.
The generator implements a relative $S$-wave ($P$-wave) configuration
between the $\pi\pi$ ($\eta$ or $\pi^0$) and
the $\jpsi$ or $\psip$. 
Detection efficiencies (Table~\ref{tab:yields})
range from 4-38\%, 
not including the effects of the intermediate 
branching fractions~\cite{PDG,psiprimexjpsiprl}
for $\jpsi\to\dilep$,
$\psip\to\dilep$, $\psip\to\dipi\jpsi$, 
$\chi_{cJ}\to\gamma\jpsi$, or $\chi_{c0}\to$hadrons,
but including those for $\pi^0$, $\eta$, $\omega$, 
$\etaprime$, and $\phi$ decay. 
Already included are the effects
of ISR, which reduce efficiencies by relative fractions
of 8-21\%.

The radiative return process $\diel\to\gamma\psip\to X\jpsi$
results in final states which are nearly identical to some of
our signal modes, and thereby affords an opportunity to verify
our understanding of efficiencies, background, and luminosity.
To gather such events, we alter only the $k$-windows of the event 
selections, as such events will congregate
not near $k=0$ but rather around $k_0=(s-M_\psip^2)/2\sqrt{s}$.
The cross-section for this process can be calculated by integrating the
convolution of a Breit-Wigner lineshape (approximated by a $\delta$-function)
with the ISR kernel~$W$ from Eq.~(28) of Ref.~\cite{ISR}:
\begin{equation}
\sigma(\diel\to\gamma\psip)= 
    \frac{12\pi^2 \Gamma_{ee}}{sM_{\psi(2S)}}
    \times     W(s,2k_0/\sqrt{s}\,).
\end{equation}
This curve as well as the cross-section measurements 
using $\psip\to X\jpsi$, $X$=$\dipi$, $\dipiz$, 
and $\eta$~\cite{psiprimexjpsiprl}, 
are shown in Figure~\ref{fig:fig1}(c).
The CLEO value for $\Gamma_{ee}[\psi(2S)]$~\cite{gammaeePsi2S}
sets the scale of the theoretical curve in the figure.
The predicted number of 
observed radiative return events for 
the three channels
together in all the scan data is 820$\pm$21, 
which compares favorably to 825$\pm$29~events seen.

\begin{figure}
\includegraphics*[width=6.5in]{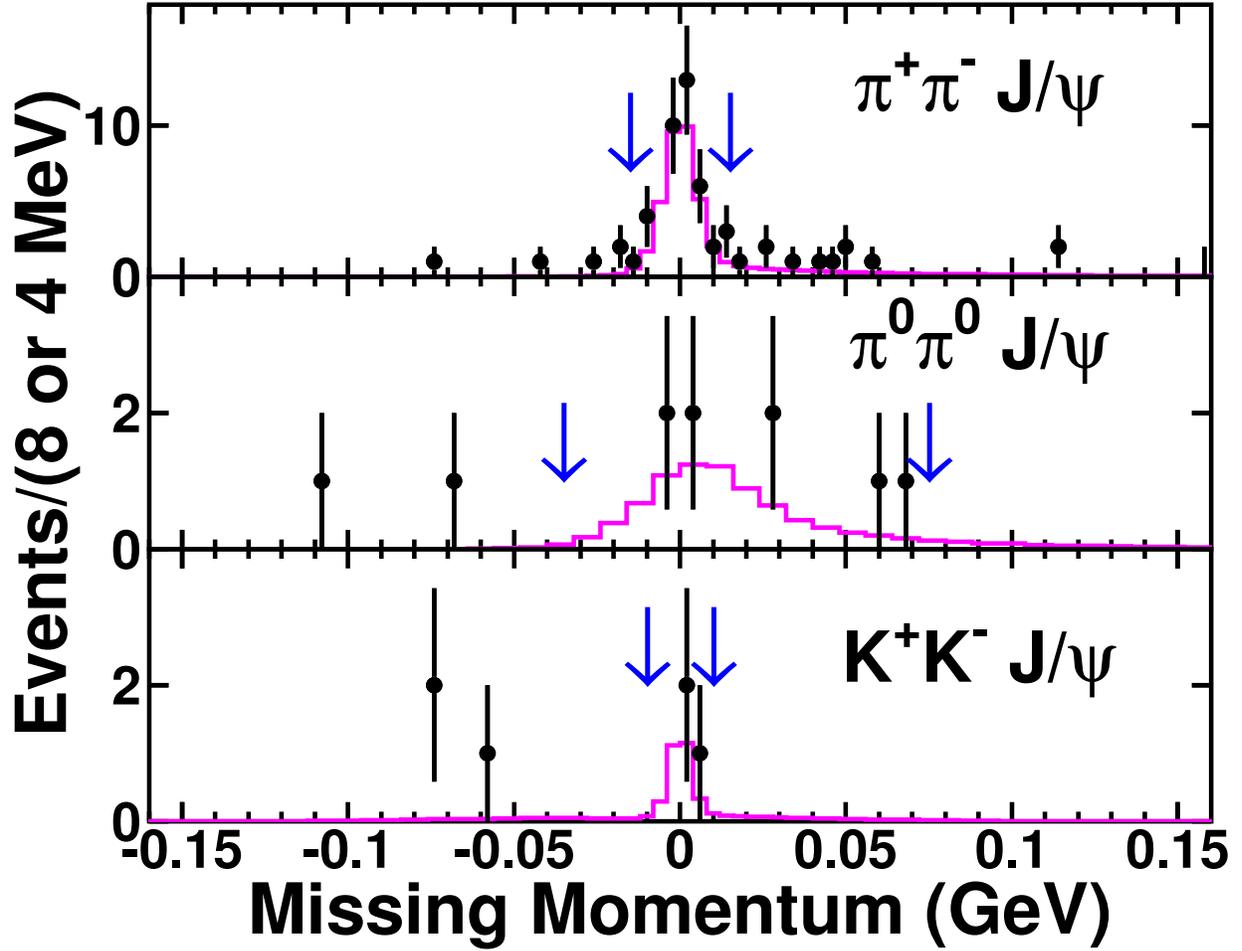}
\caption{The missing momentum ($k$) distribution for 
$\pipijpsi$ (top), $\pizpizjpsi$ (middle), and $\kkjpsi$ (bottom) in
the data at
$\sqrt s = 4.26\gev$ (circles), and the
signal shape as predicted by MC simulation
(solid line histogram) scaled to the net signal size.}
\label{fig:fig2}
\end{figure}

\begin{figure}
\includegraphics*[width=6.5in]{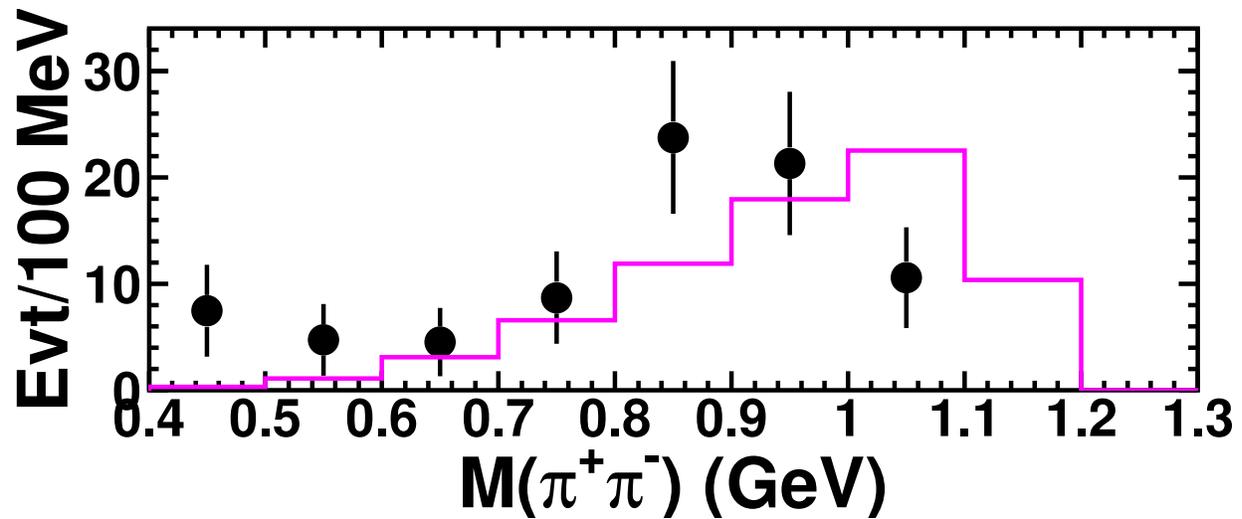}
\caption{The efficiency-corrected dipion invariant mass distribution for the 
$\pipijpsi$ final state in the
signal region at $\sqrt s = 4.26\gev$ for the data (circles)
and the signal shape (solid line histogram) as predicted by $\psi(2S)$-like MC simulation
scaled to the net signal size.}
\label{fig:fig3}
\end{figure}

Table~\ref{tab:yields} lists cross-section results for 
the $k=0$ region. 
Cross-section central values and errors are shown
when the statistical significance (the likelihood that
the observed event yield is due entirely to background)
exceeds $2.5\sigma$;
otherwise upper limits at 90\% confidence level (CL)
are shown.
Poisson fluctuations of the background are taken into
account in the computation of statistical uncertainties.
Cross-sections for $\diel\to\pi\pi\jpsi$ {\sl vs.}~$\sqrt{s}$
are shown in Fig.~\ref{fig:fig1}(d);
selected missing momentum and dipion mass distributions 
appear in Figs.~\ref{fig:fig2} and~\ref{fig:fig3}, respectively.
The yields for $\pipiphi$ are consistent with the rate
observed~\cite{multibody} from $\diel\to\gamma^*$ at $\sqrt s$=3.67~GeV.

Systematic uncertainties arise 
from the following sources: luminosity (2\%), charged particle
tracking (1\% per track), particle 
identification (1\% per high momentum $e$, $\pi$, or $K$;
5\% per $K$ in $\kkjpsi$), 3\% per $\mu$~pair
for muon chamber modeling, and a mode-dependent
contribution amounting to 
50\% of the total estimated background.
Statistical uncertainties dominate.
The total systematic error excluding the background
uncertainty amounts to approximately 10\% for most channels
(up to 30\% for $\chicJ$ modes and $\pipiphi$).
The only charmonium channels with more than 2.5$\sigma$ statistical significance are,
at $\sqrt s$=4.260$\gev$,
$\pipijpsi$   (11$\sigma$), 
$\pizpizjpsi$ (5.1$\sigma$), 
and $\kkjpsi$ (3.7$\sigma$);
in the $\psi(4160)$ dataset,
$\pipijpsi$   (3.6$\sigma$) and
$\pizpizjpsi$ (2.6$\sigma$); and
in the $\psi(4040)$ dataset,
$\pipijpsi$ (3.3$\sigma$). 
The $\psi(4160)$-region yields of $\pipijpsi$ and $\pizpizjpsi$ 
are consistent with being due entirely to
the $\yfts$ low-side tail~\cite{BaBar4260ISR}.
No compelling evidence is found 
for any other decays in the three resonance regions, and 
corresponding upper limits on cross-sections (and,
for $\psi(4040)$ and $\psi(4160)$ datasets, on 
branching fractions) are set.  
In particular, we find
${\cal B}(\psi(4040) \to \pipijpsi) <$0.4\% 
and
${\cal B}(\psi(4160) \to \pipijpsi) <$0.4\%.
These correspond to
partial widths of less than $0.4\mev$ 
in both cases, 
to be compared with the central values for
$\psi(2S)$ and $\psi(3770)$ of
$0.10\mev$ and $0.44\mev$~\cite{PDG}, respectively.
While statistics are low, no prominent narrow features emerge
in $M(\dipi)$, and the distribution is somewhat softer than
the $\psip$-like MC~prediction.

  This analysis provides a high-significance
confirmation of the {\sc BaBar} signal of $\pipijpsi$. 
The observation of the $\dipiz\jpsi$ mode
disfavors the $\chicJ\rho^0$ molecular model~\cite{LZL}. 
The fact that the $\dipiz\jpsi$ rate is about half that of 
$\dipi\jpsi$ disagrees with the prediction of
the baryonium model~\cite{Qiao}. Our evidence of 
significant $\dika\jpsi$ production is not compatible
with these two models either. No evidence of a large
$\dipi\jpsi$ signal from the $\psi(4040)$ is observed, making
the conventional $Y(4260)$=$\psi(4S)$ assignment~\cite{Estrada} 
less attractive. The results are compatible with hybrid charmonium
interpretations~\cite{ClosePage,hybrids},
but open-charm studies will be required to make
more definitive conclusions.
A large coupling to $D_1(2430)^0\overline{D}_0$ and
a small one to $D_s\overline{D_s}$ signals hybrid
charmonium~\cite{ClosePage}, whereas a dominant $D_s\overline{D_s}$ 
($D\overline{D}$) could favor a $c\bar{s}s\bar{c}$~\cite{ClosePage,Maiani} 
($c\bar{q}q\bar{c}$~\cite{Ebert}) tetraquark model.

We gratefully acknowledge the effort of the CESR staff
in providing us with excellent luminosity and running conditions.
This work was supported by
the A.P.~Sloan Foundation,
the National Science Foundation,
and the U.S. Department of Energy.

\begin{table*}[htp]
\setlength{\tabcolsep}{0.4pc}
\catcode`?=\active \def?{\kern\digitwidth}
\caption{
For each mode $\diel \to X$, for three center-of-mass regions:
the detection efficiency, $\epsilon$; 
the number of signal [background] events in data, 
$N_{\mathrm{s}}$ 
[$N_{\mathrm{b}}$];
the cross-section $\sigma(\diel \to X)$; 
and the branching fraction of $\psi(4040)$ or $\psi(4160)$ to $X$,
${\cal B}$. Upper limits are at 90\%~CL. '--' indicates that the
channel is kinematically or experimentally inaccessible.
}
\label{tab:yields}
\footnotesize
\begin{center}
\begin{tabular}{c|ccccc|ccccc|cccc}
\hline
\hline
 & \multicolumn{5}{c|}{$\sqrt s = 3970-4060\mev$} 
 & \multicolumn{5}{c|}{$\sqrt s = 4120-4200\mev$} 
 & \multicolumn{4}{c}{$\sqrt s = 4260\mev$} 
 \\ 
Channel 
& $\epsilon$ & $N_{\mathrm{s}}$ & $N_{\mathrm{b}}$ & $\sigma$ & ${\cal B}$ 
& $\epsilon$ & $N_{\mathrm{s}}$ & $N_{\mathrm{b}}$ & $\sigma$ & ${\cal B}$ 
& $\epsilon$ & $N_{\mathrm{s}}$ & $N_{\mathrm{b}}$ & $\sigma$ 
 \\ 
        
& ($\%$)       &                 &                       & (pb)   &     $(10^{-3})$ 
& ($\%$)       &                 &                       & (pb)   &     $(10^{-3})$ 
& ($\%$)       &                 &                       & (pb)        
 \\ 
\hline
 $ \pipijpsi $ 
       & 37 &  12 &  3.7 &  $    9^{+5}_{-4}$$\pm 2 $ &  $ < 4 $ 
       & 38 &  13 &  3.7 &  $    8^{+4}_{-3}$$\pm 2 $ &  $ < 4 $ 
       & 38 &  37 &  2.4 &  $   58^{+12}_{-10}$$\pm 4 $ 

 \\ 
 $ \pizpizjpsi $ 
       & 20 &   1 &  1.9 &  $ < 8 $  &  $ < 2 $ 
       & 21 &   5 &  0.9 &  $    6^{+5}_{-3}$$\pm 1 $ &  $ < 3 $ 
       & 22 &   8 &  0.3 &  $   23^{+12}_{-8}$$\pm 1 $ 

 \\ 
 $ \kkjpsi $ 
       & \multicolumn{5}{c|}{--} 
       & 7 &   1 &  0.07 &  $ < 20 $  &  $ < 5 $ 
       & 21 &   3 &  0.07 &  $    9^{+9}_{-5}$$\pm 1 $ 

 \\ 
 $ \etajpsi $ 
       & 19 &  12 &  9.5 &  $ < 29 $  &  $ < 7 $ 
       & 16 &  15 &  8.8 &  $ < 34 $  &  $ < 8 $ 
       & 16 &   5 &  2.7 &  $ < 32 $  

 \\ 
 $ \pizjpsi $ 
       & 23 &   2 &         &  $ < 10 $  &  $ < 2 $ 
       & 22 &   1 &         &  $ < 6 $  &  $ < 1 $ 
       & 22 &   1 &         &  $ < 12 $

 \\ 
 $ \etaprimejpsi $ 
       & \multicolumn{5}{c|}{--} 
       & 11 &   4 &  2.5 &  $ < 23 $  &  $ < 5 $ 
       & 11 &   0 &  1.5 &  $ < 19 $  

 \\ 
 $ \threepijpsi $ 
       & 21 &   1 &         &  $ < 8 $  &  $ < 2 $ 
       & 21 &   0 &         &  $ < 4 $  &  $ < 1 $ 
       & 22 &   0 &         &  $ < 7 $  

 \\ 
 $ \etaetajpsi $ 
       & \multicolumn{5}{c|}{--} 
       & \multicolumn{5}{c|}{--} 
       & 6 &   1 &         &  $ < 44 $  

 \\ 
 $ \pipipsiprime $ 
       & \multicolumn{5}{c|}{--} 
       & 12 &   0 &         &  $ < 15 $  &  $ < 4 $ 
       & 19 &   0 &         &  $ < 20 $  

 \\ 
 $ \etapsiprime $ 
       & \multicolumn{5}{c|}{--} 
       & \multicolumn{5}{c|}{--} 
       & 15 &   0 &         &  $ < 25 $  

 \\ 
 $ \omegachiczero $ 
       & \multicolumn{5}{c|}{--} 
       & \multicolumn{5}{c|}{--} 
       & 9 &  11 & 11.5 &  $ < 234 $  

 \\ 
 $ \gammachicone $ 
       & 26 &   9 &  8.1 &  $ < 50 $  &  $ < 11 $ 
       & 26 &  11 &  8.7 &  $ < 45 $  &  $ < 10 $ 
       & 26 &   1 &  3.3 &  $ < 30 $  

 \\ 
 $ \gammachictwo $ 
       & 25 &   6 &  8.0 &  $ < 76 $  &  $ < 17 $ 
       & 26 &  10 &  8.6 &  $ < 79 $  &  $ < 18 $ 
       & 27 &   4 &  3.3 &  $ < 90 $  

 \\ 
 $ \threepichicone $ 
       & 6 &   0 &         &  $ < 47 $  &  $ < 11 $ 
       & 8 &   0 &         &  $ < 26 $  &  $ < 6 $ 
       & 9 &   0 &         &  $ < 46 $  

 \\ 
 $ \threepichictwo $ 
       & 4 &   0 &         &  $ < 141 $  &  $ < 32 $ 
       & 8 &   0 &         &  $ < 56 $  &  $ < 13 $ 
       & 9 &   0 &         &  $ < 96 $  

 \\ 
 $ \pipiphi $ 
       & 17 &  26 &  3.0 &  $ < 12 $  &  $ < 3 $ 
       & 17 &  17 &  6.0 &  $ < 5 $  &  $ < 1 $ 
       & 18 &   7 &  5.5 &  $ < 5 $  

 \\ 
\hline
\hline
\end{tabular} 
\end{center}
\end{table*}
\clearpage

\end{document}